\documentclass[twoside,a4paper,11pt]{sea10}
\usepackage{graphicx}
\usepackage{hyperref}
\usepackage{lscape}

\newcommand{\vsini} {$v$\,sin\,$i$}

\newcommand{\vmacro} {$v_{\rm mac}$}

\newcommand{\Teff} {$T_{\rm eff}$}
\newcommand{\grav} {log\,{\em $g$}}

\newcommand{\micro} {$\xi_{\rm t}$ }

\newcommand{\abun}[1]{{$\epsilon_{\rm Si}$}}

\begin{document}
\pagenumbering{arabic}
\pagestyle{myheadings}
\thispagestyle{empty}
{\flushleft\includegraphics[width=\textwidth,bb=58 650 590 680]{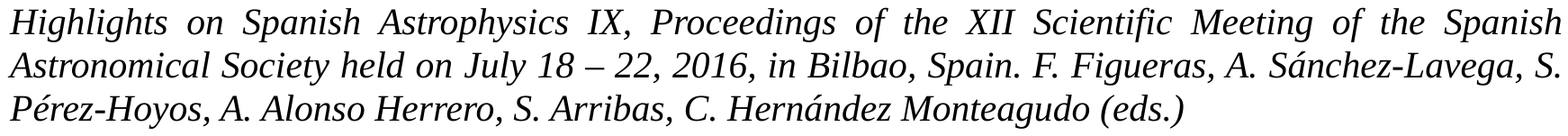}}
\vspace*{0.2cm}
\begin{flushleft}
{\bf {\LARGE
%
Physical characterization of Galactic O-type stars targeted by the IACOB and OWN surveys
%
}\\
\vspace*{1cm}
%
Gonzalo Holgado$^{1,2}$,
Sergio Sim\'on-D\'{\i}az$^{1,2}$, 
and 
Rodolfo Barb\'a$^{3}$
%
}\\
\vspace*{0.5cm}
%
$^{1}$
Instituto de Astrof\'isica de Canarias, E-38200 La Laguna, Tenerife, Spain. \\
$^{2}$
Departamento de Astrof\'isica, Univ. de La Laguna, E-38205 La Laguna, Tenerife, Spain.\\
$^{3}$
Departamento de F\'isica y Astronom\'ia, Univ. de la Serena, Av. Juan Cisternas 1200 Norte, La Serena, Chile\\
%
\end{flushleft}
%
\markboth{
Physical characterization of Galactic O-type stars from IACOB and OWN
}{ 
%
Holgado et al. 
%
}
\thispagestyle{empty}
\vspace*{0.4cm}
\begin{minipage}[l]{0.09\textwidth}
\ 
\end{minipage}
\begin{minipage}[r]{0.9\textwidth}
\vspace{1cm}
\section*{Abstract}{\small
%

We present first results from the quantitative spectroscopic analysis of $\sim$270 Galactic O-type stars 
targeted by the IACOB and OWN surveys (implying the largest sample of stars of this type analyzed 
homogeneously). We also evaluate what is the present situation regarding available information about 
distances, as provided by the Hipparcos and Gaia missions.
%
\normalsize}
\end{minipage}
%
%
%
\section{Introduction \label{intro}}

We are immersed in the era of the large spectroscopic surveys of massive O- and B-type stars. An era which is
smoothly overlapping with the time in which the Gaia mission is expected to provide fresh momentum to the 
field of stellar astrophysics.

Based on the exhaustive Galactic O-star Catalogue (\cite{Maiz13}), the IACOB (\cite{Simon11a, Simon11b, Simon15}) and OWN (\cite{Barba10,Barba14}) projects have independently devoted an enormous observational effort in the last decade to compile two large multi-epoch high-resolution spectroscopic databases of Galactic O-type stars, including Northern and Southern targets, respectively. A few years ago, we decided to joint efforts and start the exploitation of this unique spectroscopic dataset regarding the physical characterization of the stellar properties of the complete sample. To successfully reach our objective we are benefiting from a battery of self-made automatized tools optimized for the quantitative spectroscopic analysis of large samples of O-type stars, and plan to use the upcoming information about distances provided by the ESA-Gaia mission (\cite{Perryman01}). In this proceeding we introduce the main sample which is being analyzed and present first results of our on-going work.

\section{Sample}\label{section2}

We used the Galactic O-star Catalogue (GOSC, P.I., J. Ma\'iz Apell\'aniz) as initial reference to build our sample of stars to be analyzed spectroscopically. The GOSC has presently become the most complete and updated compilation of accurate information related to this type of objects. Its current public version comprises 601 Galactic stars with spectral types in the range O2\,--\,09.7 (all luminosity classes) from both -- Northern and Southern -- hemispheres (see Ma\'iz Apell\'aniz, these proceedings).

We searched in the two modern multi-epoch, high resolution spectroscopic databases IACOB\footnote{Part of these observations correspond to the IACOBsweg program \cite{Simon15} led by I.Negueruela} (P.I., S. Sim\'on-D\'{\i}az), and OWN  (P.Is., R. Barb\'a \& R. Gamen), and found a total of 2452 spectra for 340 of the stars listed in GOSC. In particular, we concentrated on the spectra collected by these surveys with three different instruments: FIES, HERMES, and FEROS. The similarity between these spectroscopic observations -- in terms of resolving power and S/N -- allows to minimize observational effects on the outcome from our spectroscopic study. A global overview of the number of stars and spectra available from each survey is presented in Table \ref{tablesurveys}. 

As a natural consequence of the observing philosophy of the IACOB and OWN surveys, most of the stars count on more than 2 spectra obtained at different epochs. Only the spectrum with the best S/N was considered for the quantitative spectroscopic analysis leading to the stellar parameters, but we used the complete spectroscopic dataset\footnote{In this case, also including available spectra from the CAFE-BEANS survey \cite{Negueruela15}} to obtain information about spectroscopic binarity or variability

\begin{figure}
	\center
	\includegraphics[scale=0.5]{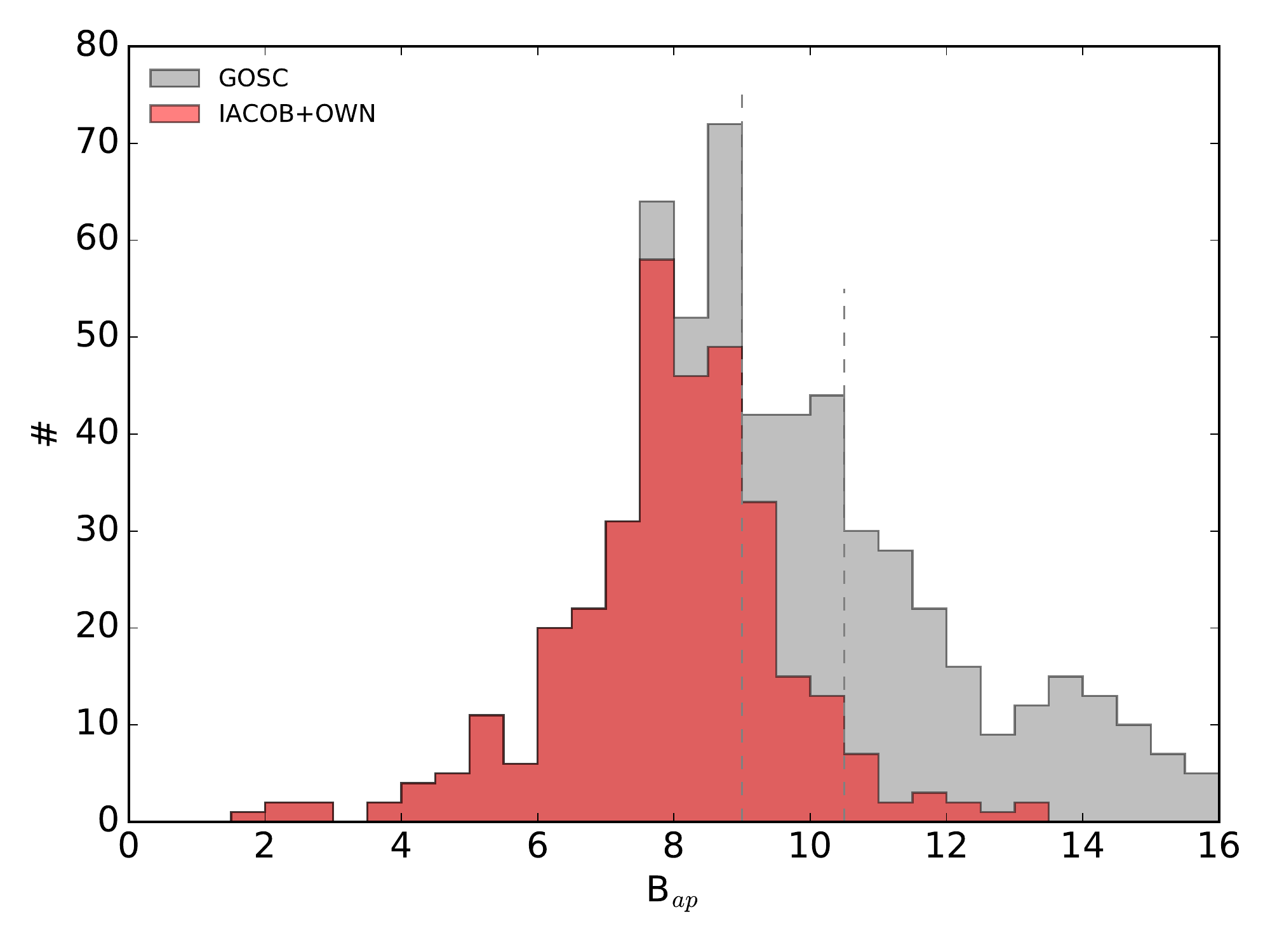} ~
	\caption{\label{fig1} B$_{\rm ap}$ magnitude histogram the comparing the GOSC and IACOB+OWN samples.}
\end{figure}

The initial sample of stars with high resolution spectra represents $\sim$56\% of the GOSC sample. This is due to the more limited range in magnitude characterizing the stars\footnote{An extension of the IACOB and OWN observations with the objective of being complete up to V=11 is already on-going.} targeted by IACOB and OWN (see Fig.~\ref{fig1}). We note, however, that if we just concentrate in stars brighter than B$_{ap}$=9 (where IACOB and OWN have been mainly concentrated to date) the high resolution sample is complete up to $\sim$90\%. From this initial list we excluded those stars clearly detected as double line spectroscopic binaries (SB2). In the following, we concentrate on 266 O-type stars classified as likely single or single line spectroscopic binaries (SB1).

\begin{table}[t!] 
	\caption{Summary of the spectroscopic observations used in this work} 
	\begin{minipage}{0.5\textwidth}
		\center
	\begin{tabular}{l l l c c c }
		\hline\hline
		\noalign{\smallskip}
		Survey     & Telescope & Instrument & Resolution     & Range [\AA] & \# Spectra \\ 
		\hline
		\noalign{\smallskip}
		IACOB      & NOT-2.56m       & FIES       & 46000 \& 25000 & 3704-7271  & 828     \\
		           & MERCATOR-1.2m  & HERMES     & 85000          & 3763-9006  & 658     \\
		OWN        & ESO-2.2m    & FEROS      & 46000          & 3527-9217  & 966     \\
		\hline
		\noalign{\smallskip}
		TOTAL     &             &            &                &            &  2452*   \\ 
		\hline
		\noalign{\smallskip}
		\multicolumn{6}{l}{* 340 stars eliminating multi-epoch and stars in common}
		
	\end{tabular}
	\end{minipage}
	\label{tablesurveys} 
\end{table}


\section{Quantitative spectroscopic analysis}\label{section3}

We performed the quantitative spectroscopic analysis of our final list of 266 Galactic O-type stars using several tools developed in the framework of the IACOB project. First, \textsc{iacob-broad} \cite{Simon14} was used to obtain the projected rotational velocity (\vsini) and the amount of non-rotational broadening (aka. macroturbulence, \vmacro) affecting the line-profiles of each star (i.e. the {\em line-broadening parameters}). Then, \textsc{iacob-gbat} \cite{Simon11}, a semi-automatized tool for the quantitative spectroscopic analysis of O-stars based on an extensive grid of \textsc{fastwind} \cite{Puls05} models and an optimized $\chi^2$ strategy, was used to obtain the so-called {\em spectroscopic parameters}; namely, the effective temperature (\Teff), gravity (\grav), wind-strength parameter ($Q$), helium abundance ($Y_{\rm He}$), microturbulence (\micro), and the exponent of the wind velocity-law ($\beta$). We refer the reader to \cite{Simon14}, \cite{Simon11}, and \cite{SabinSan14} for a thorough description of the tools and some of the basic ideas about the analysis strategy and the interpretation of results. 

The information extracted from the spectra will be complemented with that for the other {\em fundamental parameters} -- such as the radius ($R$), the luminosity ($L$), or the spectroscopic mass ($M_{\rm sp}$) -- once reliable and accurate information about distances will be made available by Gaia ( see also Sect.~\ref{distances}).

\section{Results}\label{section4}
We present in this section an overview of the first results obtained from the quantitative spectroscopic analysis of  
the considered sample of Galactic stars. As indicated above, this refers to 266 O-type stars not detected as double line spectroscopic binaries from inspection of the available FIES, HERMES and FEROS from the IACOB and OWN surveys. This implies the larger sample of Galactic O-type stars spectroscopically analyzed to date in a homogeneous way.

\subsection*{Projected rotational velocities}\label{vsini}

\begin{figure}
	\includegraphics[scale=0.6]{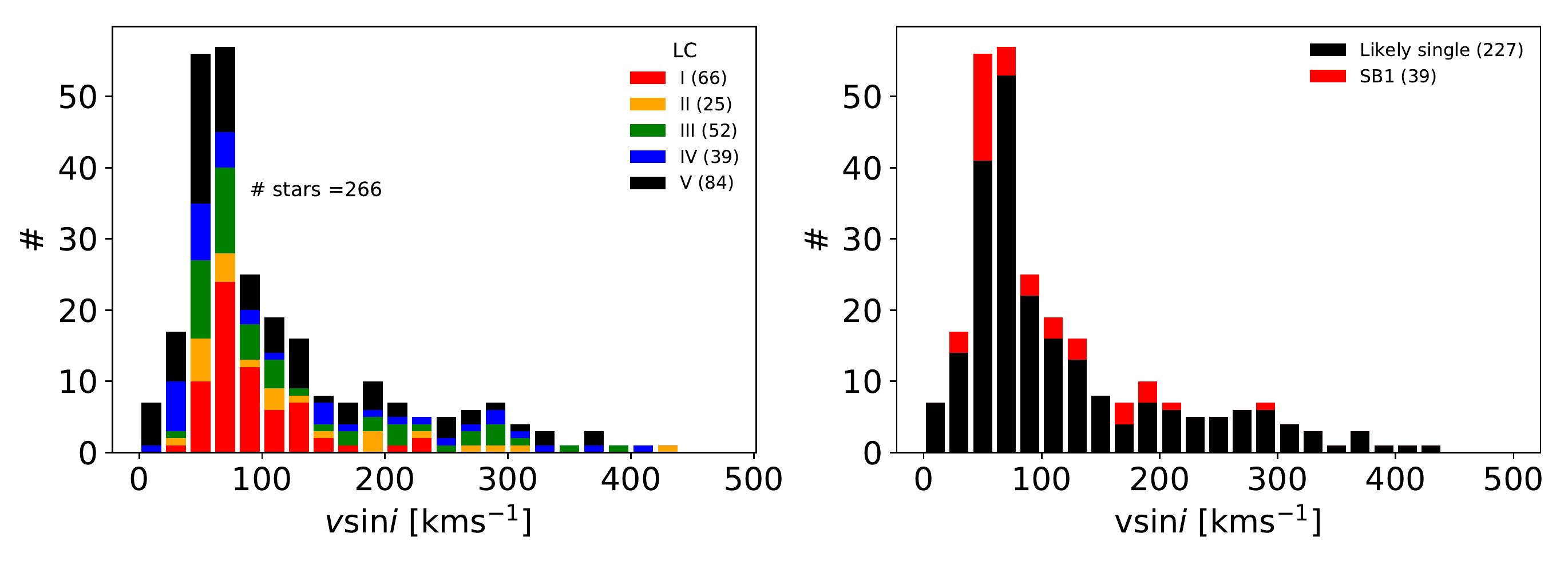} ~
	\caption{\label{fig2} Distribution of projected rotational velocities for a sample of 266 O-type stars, separated by {(\it left)} luminosity class, {(\it right)} detected binarity status. We note that results for $\sim$75 SB2 stars have been removed from this graph.
	}
\end{figure}

In Fig. \ref{fig2} we show the distribution of the projected rotational velocities in the analyzed sample of O-type stars, separated by luminosity class (left) and binary status (right). A thorough study of these distributions will allow us to step forward in our knowledge about how this important stellar parameter is initially distributed and is modified as the star evolves and/or interact with other components in a binary system. Also, by comparing this result with that obtained by, e.g., \cite{RamAgu13}, we will be able to investigate the effect of metallicity on the evolution of spins rates in massive stars.

\subsection*{Stellar properties in the spectroscopic HR diagram}\label{specparam}

\begin{figure}
	\center
	\includegraphics[scale=0.75]{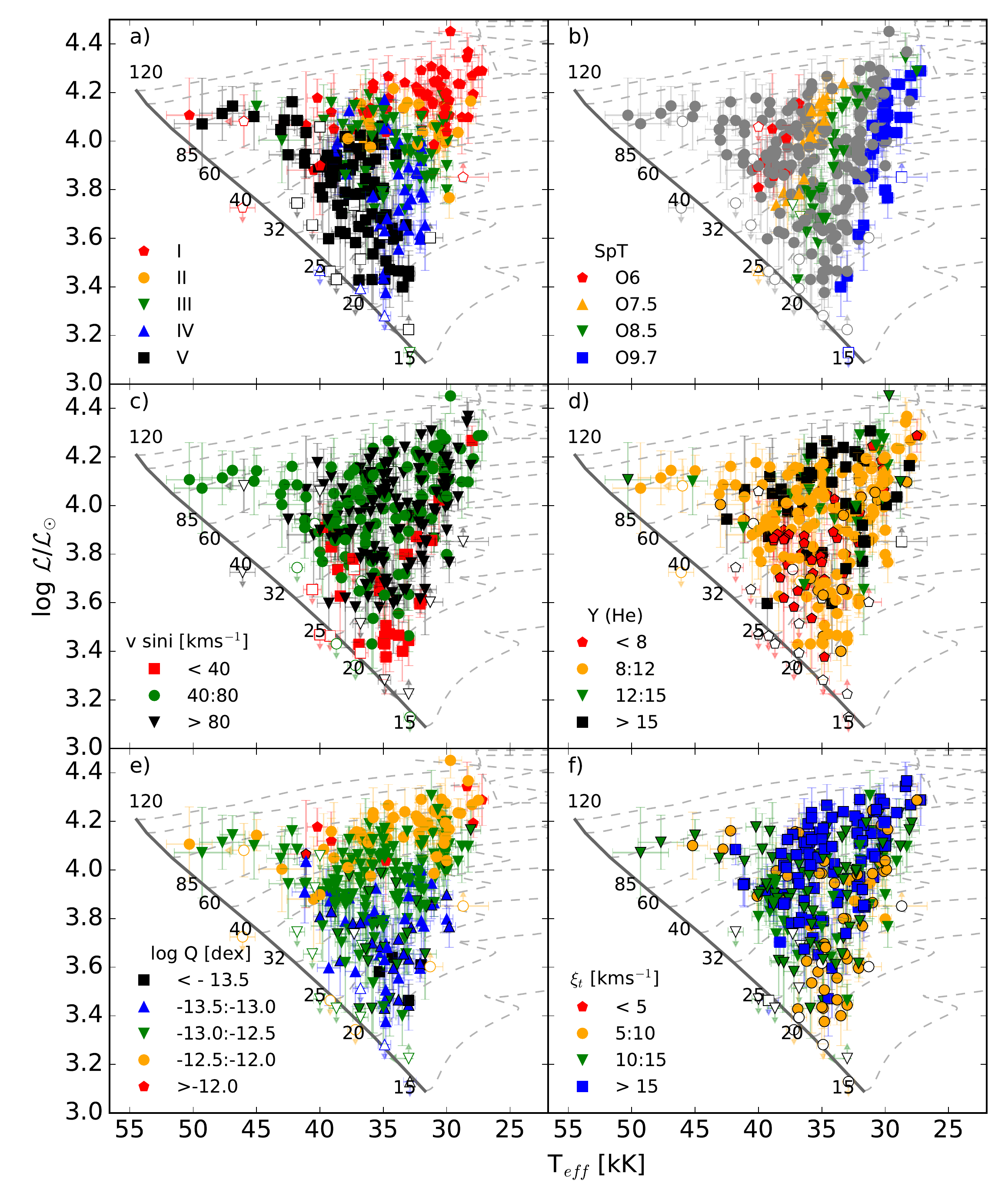} ~
	\caption{\label{fig3} Distribution in the spectroscopic HR diagram of several stellar properties determined spectroscopically in our sample of 266 likely single and/or SB1 O-type stars. Open points are limits. Non rotating evolutionary tracks (and ZAMS) from the Geneva group (\cite{Ekstrom12} and \cite{Georgy13}) overplotted for reference.
	}
\end{figure}

In Fig. \ref{fig3} we present how luminosity class, spectral type, as well as four of the stellar parameters determined spectroscopically (\vsini, $Y(He)$, log $Q$, and \micro) are distributed in the spectroscopic HR diagram \cite{Langer14}. 
This figure shows the good coverage of the classic O-type domain we have reached with the compiled and analyzed sample of stars. We are confident that our study (once finalized by incorporating information about radii, luminosities, spectroscopic masses and abundances) will provide an unprecedented empirical overview of the main physical properties of Galactic massive O-type stars to be used as definitive anchor point for our theories of stellar atmospheres, winds, interiors and evolution of massive stars.

\section{Are we ready to incorporate information about distances to our study?}\label{distances}

As indicated above, our final objective is the complete characterization of the whole sample of Galactic O-type stars targeted by IACOB and OWN, also including information about abundances, as well as the remaining fundamental parameters: $L$, $R$, and $M_{\rm sp}$. For the later, accurate distances are needed. 

\begin{figure}
	\center
	\includegraphics[scale=0.5]{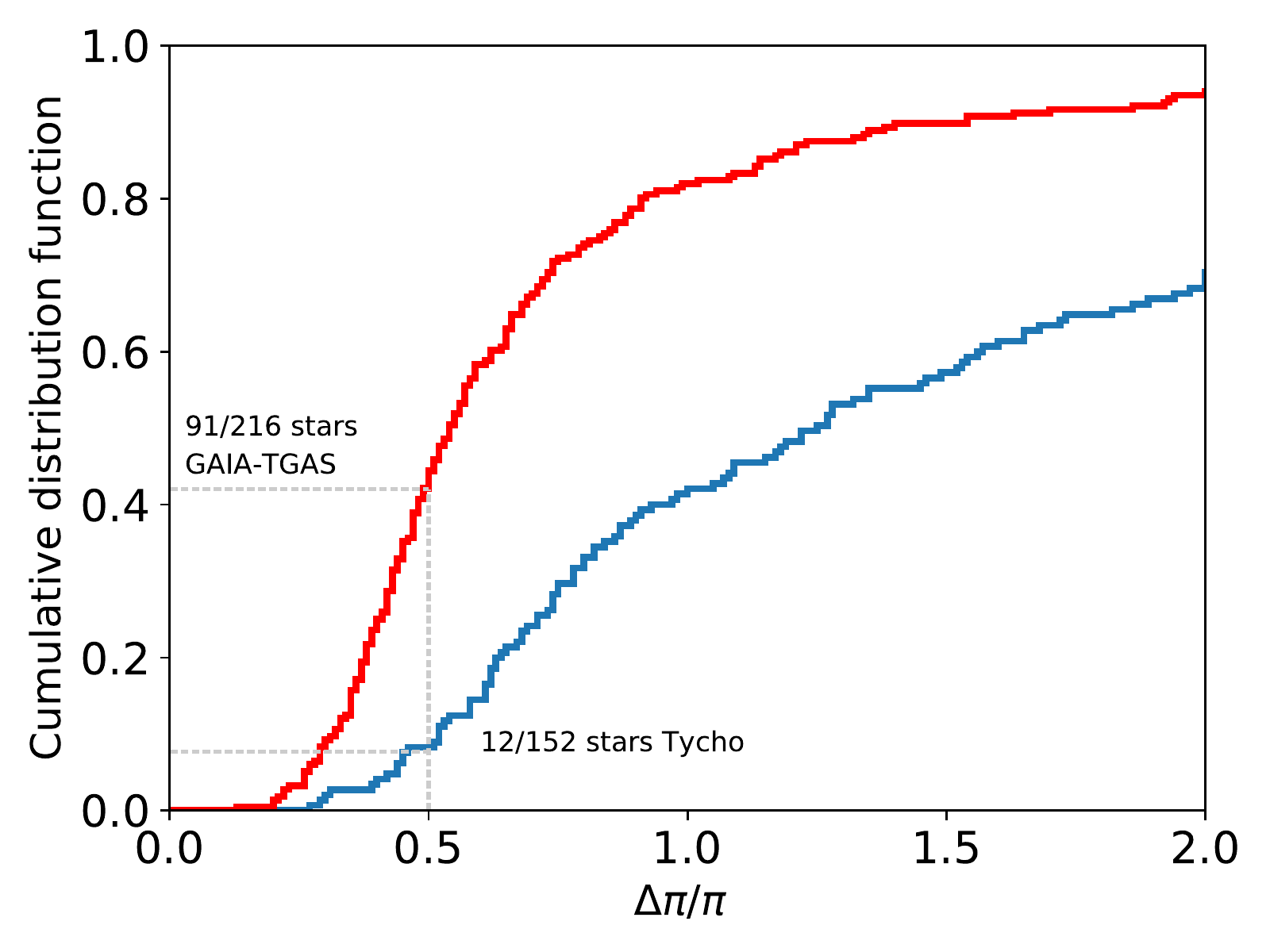} ~
	\caption{\label{fig4} Normalized cumulative distribution function of the relative error in TYCHO and GAIA-TGAS parallaxes for our sample  of 340 Galactic O-type stars. Negative parallaxes have been eliminated.
	}
\end{figure}

Where are we in terms of distances? To answer this question, we searched for available information about parallaxes for our initial sample of 340 stars. We collected 152 and 216 parallaxes with positive values in the TYCHO catalog \cite{Hoeg97} and the recently released GAIA-TGAS DR1 \cite{Lindegren16}, respectively. The compiled information is summarized in Fig.~\ref{fig4} as two normalized cumulative distribution functions of the relative error resulting from each dataset. Overall, there is a clear improvement of the situation from TYCHO to TGAS, with a higher number of parallaxes with positive values and a decrease in the mean relative error; however, we have not reach yet a situation good enough for the purposes of our work since still only $\sim$ 40\% of the sample present a formal relative error (as provided in TGAS) below 50\%.

We will hence have to wait for GAIA DR2, when an important improvement in the parallax determination and the associated uncertainties is expected, to be able to compute reliable values for the physical parameters of our sample.

%
%
\small  
%
%

%

%
\end{document}